\documentclass[aps,twocolumn,pra,superscriptaddress,amsmath,showpacs,tightenlines]{revtex4-1}
\usepackage{amssymb}
\usepackage{amsmath}
\usepackage{graphicx}
\usepackage{subfigure}
\usepackage{natbib}
\usepackage{epsfig}
\usepackage{amsfonts}
\usepackage{mathrsfs}
\usepackage{xcolor}
\usepackage[toc,page,title,titletoc,header]{appendix}

\begin{document}

\title{Effect of the short-range interaction on low-energy collision of ultracold dipoles}

\author{Peng Zhang}

\email{pengzhang@ruc.edu.cn}
\affiliation{Department of Physics, Renmin University of China, Beijing, 100190,
China}

\affiliation{Beijing Computational Science Research Center, Beijing, 100084, China}

\affiliation{Beijing Key Laboratory of Opto-electronic Functional Materials \&
Micro-nano Devices (Renmin University of China)}

\author{Jianwen Jie}

\affiliation{Department of Physics, Renmin University of China, Beijing, 100190,
China}

\affiliation{Beijing Key Laboratory of Opto-electronic Functional Materials \&
Micro-nano Devices (Renmin University of China)}
\begin{abstract}
We consider the low-energy scattering of two ultracold polarized dipoles with
both a short-range interaction (SRI) and a weak dipole-dipole interaction
(DDI) which is far away from shape-resonances. In previous analytical studies,
the scattering amplitude in this system was often calculated via the first-order Born
approximation (FBA). Our results show that significant derivations from this
approximation can arise in some cases. In these cases, the SRI can significantly modify the inter-dipole
scattering amplitudes even if the scattering amplitudes for the SRI
alone are much smaller than the dipolar length of the DDI. We
further obtain approximate analytical expressions for these inter-dipole
scattering amplitudes. 

\end{abstract}

\maketitle

\section{introduction}

During the past decade, fast progresses have been made in the
experimental study of complicated magnetic atoms \cite{crbecexp1,crbecexp2,crbecexp3,crbecexp4,crbecexp5,dybecexp,dyfermiexp,erbecexp,erfermiexp,erexp3,erexp4,erexp5}
and polar molecules \cite{KRbhighdensityexp,KRbfermiexp,KRbreactionexp}.
In an ultrcold gas of these atoms or molecules, the interactions between
particles include both a long-range dipole-dipole interaction (DDI)
and a short-range interaction (SRI) (e.g., van der Waals interaction).
The interplay between these two types of interactions can usually
lead to intriguing quantum phenomena at both two-body and many-body
levels. A well-known example is that, when the DDI is near a shape
resonance \cite{shaperesonanceyou}, a small hard-core SRI can significantly
shift the location of this resonance \cite{shaperesonanceblume,shaperesonancehui}.
In this paper we consider the opposite case where the DDI is away
from shape resonances, and furthermore the low-energy scattering amplitude for
the SRI is much smaller than the characteristic length of the DDI
(``dipole length'') . 
The scattering amplitude of atoms in this system can be 
calculated with exact numerical calculation \cite{shaperesonanceyou,bohnnjp}. Nevertheless,
in many previous analytical studies in this subject, 
the contribution of the DDI to the low-energy scattering amplitude is often calculated
with the first-order Born approximation (FBA), under which the interplay
between the SRI and the DDI is ignored \cite{yiandyou,BA2,BA3,BA4,BA5,daweiwang,bohnnjp,bohnxxx}.

In our investigation we calculate the low-energy scattering amplitude with the distorted-wave 
Born approximation \cite{DWBA1,DWBA2,taylor} (DWBA), which is
the accurate first-order approximation of the DDI. In the expression
of the scattering amplitude given by the DWBA, there is one term which describes
the DDI-SRI interplay (i.e., the SRI-induced effect for the inter-dipole
scattering). This term is ignored by the FBA. We find that for the scattering of two ultracold\emph{
}bosonic\emph{ }dipoles, this term is negligible and the FBA
works very well, provided that the $s$-wave scattering
length of the SRI is much smaller than the dipolar length and the
effects of the SRI in the high partial-wave channels can be neglected.
Nevertheless, for the scattering of two\emph{ }fermionic dipoles,
when the $p$\emph{-wave scattering volume} of the SRI is large enough,
the SRI-DDI interplay can be very important even if the scattering amplitude
for the SRI itself is much smaller than the dipolar length. In such
system the SRI-DDI interplay can significantly modify the low-energy scattering
amplitude. Furthermore, this effect can occur for arbitrary small
dipole moment. We also find a similar effect in the systems where the SRI is
strong in the $d$-wave channel. The analytical expressions for
the SRI-modified scattering amplitudes are obtained for all of these
systems.

Our results are helpful for the study of ultracold dipolar gases.
In particular, for the theoretical research on ultracold dipolar fermions,
our analytical expression for the modified scattering amplitude is
useful for the construction of a correct effective two-body interaction.
Moreover, our results also reveal the possibility of controlling DDI
via a weak SRI.

This paper is organized as follows. In Sec. II, we derive the scattering
amplitude of two polarized dipoles with the DWBA. With the help of this
result, we discuss the SRI-induced effects for the inter-dipole scatterings
in ultracold dipolar bosonic gases with weak SRI and ultracold fermionic
dipolar gases in Secs. III and IV, respectively. In Sec. V we consider
ultracold bosonic dipoles with a SRI which has strong effects in the
$d$-wave channel. In Sec. VI there are some conclusions and discussions.
We describe some details of our calculations in the appendixes.

\section{scattering amplitude given by the DWBA}

\subsection{Scattering amplitude}

We consider the scattering of two dipoles polarized along the $z$-direction.
The Hamiltonian for the relative motion is given by 
\begin{eqnarray}
H=\frac{{\bf p}^{2}}{2\mu}+V_{\rm sr}(r)+V_{d}({\bf r}),
\end{eqnarray}
where ${\bf p}$ is the inter-dipole relative momentum, ${\bf r}$ is
the relative coordinate, $\mu$ is the reduced mass, while $V_{\rm sr}$
and $V_{d}$ are potentials for SRI and DDI, respectively. For simplicity,
we assume that $V_{\rm sr}$ is isotropic. The DDI potential $V_{d}({\bf r})$
is given by ($2\mu=\hbar=1$) 
\begin{eqnarray}
V_{d}({\bf r}) & = & 2a_{d}\frac{(1-3\cos^{2}\theta_{{\bf r}})}{r^{3}},\label{eq:-9}
\end{eqnarray}
where $\theta_{{\bf r}}$ is the angle between ${\bf r}$ and the
$z$-axis (polar angle of ${\bf r}$), and $a_{d}=d_{1}d_{2}/2$ is
the dipolar length, with $d_{j}$ the dipole moment of the $j$-th
dipole.

In this paper we consider the systems
where the dipole moments are weak enough, so that in the calculations of the scattering amplitudes
we only need to keep terms up to
the first order of $V_{d}$. On the other hand, in many systems the intensities of the SRI are very strong  (for instance, the depth of the SRI between ultracold
atoms can be as large as $10^{14}$Hz \cite{rmp}). As a result, the SRI cannot be
treated as a perturbation. Thus, in our calculation we treat $V_{\rm sr}$
non-peturbatively. Such treatment is the DWBA \cite{DWBA1,DWBA2,taylor}.
The scattering amplitude given by the DWBA can
be expressed as \cite{DWBA2,taylor} 
\begin{equation}
f({\bf k}_{i},{\bf k}_{f})=f_{\rm sr}({\bf k}_{i},{\bf k}_{f})-2\pi^{2}\langle s_{{\bf k}_{f}}^{(-)}|V_{d}|s_{{\bf k}_{i}}^{(+)}\rangle,\label{f2}
\end{equation}
where ${\bf k}_{i}$ and ${\bf k}_{f}$ are incident and outgoing
momenta, respectively. Due to energy conservation, they satisfy $|{\bf k}_{i}|=|{\bf k}_{f}|=k$.
Here $f_{\rm sr}({\bf k}_{i},{\bf k}_{f})$ is the exact scattering amplitude
for $V_{\rm sr}$, while $|s_{{\bf k}}^{(+)}\rangle$ and $|s_{{\bf k}}^{(-)}\rangle$
are the outgoing and incoming scattering states for $V_{\rm sr}$, respectively.

To investigate the SRI-DDI interplay, we re-express the scattering amplitude given by the DWBA as 
\begin{equation}
f({\bf k}_{i},{\bf k}_{f})=f_{\rm sr}({\bf k}_{i},{\bf k}_{f})-2\pi^{2}\langle{\bf k}_{f}|V_{d}|{\bf k}_{i}\rangle+g({\bf k}_{i},{\bf k}_{f}),\label{fd}
\end{equation}
where $|{\bf k}\rangle$ is the plane-wave eigenstate of the relative
momentum operator with eigen-value ${\bf k}$. Here the amplitude
$-2\pi^{2}\langle{\bf k}_{f}|V_{d}|{\bf k}_{i}\rangle$ is contributed
by the matrix element of the dipole-dipole interaction in the plane-wave
basis. The remaining term 
\begin{equation}
g({\bf k}_{i},{\bf k}_{f})=-2\pi^{2}\left(\langle s_{{\bf k}_{f}}^{(-)}|V_{d}|s_{{\bf k}_{i}}^{(+)}\rangle-\langle{\bf k}_{f}|V_{d}|{\bf k}_{i}\rangle\right)\label{gd}
\end{equation}
is given by the difference between the scattering state $|s_{{\bf k}}^{(\pm)}\rangle$
for $V_{\rm sr}$ and the plane wave $|{\bf k}\rangle$. This term
describes the SRI-DDI interplay or the SRI-induced effect for dipole-dipole scattering.

In the FBA, which is widely used in the
research on ultracold gases with weak dipolar interactions, the 
amplitude $g({\bf k}_{i},{\bf k}_{f})$ is neglected. As a result,
the scattering amplitude is approximated as 
\begin{eqnarray}
f_{\rm FBA}({\bf k}_{i},{\bf k}_{f})=f_{\rm sr}({\bf k}_{i},{\bf k}_{f})-2\pi^{2}\langle{\bf k}_{f}|V_{d}|{\bf k}_{i}\rangle.
\end{eqnarray}
In this paper, we carefully investigate the importance and behavior
of $g({\bf k}_{i},{\bf k}_{f})$, and the validity of the FBA.


\subsection{Partial-wave expansion}

To understand the problem more clearly, we perform partial-wave expansion
for the total scattering amplitude $f({\bf k}_{i},{\bf k}_{f})$ in
Eq. (\ref{fd}): 
\begin{eqnarray}
f({\bf k}_{i},{\bf k}_{f}) & = & 4\pi\sum_{l,l',m}F_{l,l'}^{(m)}(k)Y_{l'}^{m}(\hat{{\bf k}}_{f})Y_{l}^{m}(\hat{{\bf k}}_{i})^{*},\label{fp}
\end{eqnarray}
where $\hat{{\bf k}}={\bf k}/|{\bf k}|$ is the unit vector along
the direction of ${\bf k}$, and $Y_{l}^{m}$ is the spherical harmonic
of degree $l$ and order $m$. Here $F_{l,l'}^{(m)}(k)$ is the generalized
partial-wave scattering amplitude, and can be expressed as 
\begin{equation}
F_{l,l'}^{(m)}(k)=F_{l}^{\rm (sr)}(k)\delta_{l,l'}+P_{l,l'}^{(m)}+G_{l,l'}^{(m)}(k),\label{fp-1}
\end{equation}
with $F_{l}^{\rm (sr)}(k)$ the $l$-th partial-wave scattering amplitude
for $V_{\rm sr}$. In Eq. (\ref{fp-1}) the $k$-independent parameter
$P_{l,l'}^{(m)}$ and the function $G_{l,l'}^{(m)}(k)$ are the partial-wave
components of $-2\pi^{2}\langle{\bf k}_{f}|V_{d}|{\bf k}_{i}\rangle$
and $g({\bf k}_{i},{\bf k}_{f})$, respectively, and satisfy the relations
\begin{equation}
-2\pi^{2}\langle{\bf k}_{f}|V_{d}|{\bf k}_{i}\rangle=4\pi\sum_{l,l',m}P_{l,l'}^{(m)}Y_{l'}^{m}(\hat{{\bf k}}_{f})Y_{l}^{m}(\hat{{\bf k}}_{i})^{*}\label{eq:-8}
\end{equation}
and 
\begin{equation}
g({\bf k}_{i},{\bf k}_{f})=4\pi\sum_{l,l',m}G_{l,l'}^{(m)}(k)Y_{l'}^{m}(\hat{{\bf k}}_{f})Y_{l}^{m}(\hat{{\bf k}}_{i})^{*}.\label{gg-1}
\end{equation}

It is clear that, the generalized partial-wave scattering amplitude
given by FBA is 
\begin{equation}
F_{{\rm FBA}(l,l')}^{(m)}(k)=F_{l}^{\rm (sr)}(k)\delta_{l,l'}+P_{l,l'}^{(m)}.\label{fborn-1}
\end{equation}
Comparing Eq. (\ref{fborn-1}) to Eq. (\ref{fp-1}), we know that the SRI-induced effect or SRI-DDI interplay for the partial-wave
scattering is described by the amplitude $G_{l,l'}^{(m)}(k)$.

In the following sections, we investigate the magnitude of $G_{l,l'}^{(m)}(k)$
for ultracold gases of polarized bosonic and fermionic dipoles. To
this end, we first re-express $P_{l,l'}^{(m)}$ and $G_{l,l'}^{(m)}(k)$ as
\begin{eqnarray}
 P_{l,l'}^{(m)}=a_{d}D_{l,l'}^{(m)}\int_{0}^{\infty}\frac{1}{r}j_{l}(kr)j_{l'}(kr)dr\label{gllp}
\end{eqnarray}
and
\begin{eqnarray}
 &  & G_{l,l'}^{(m)}(k)=a_{d}D_{l,l'}^{(m)}\times\nonumber \\
 &  & \int_{0}^{\infty}\frac{1}{r}[\phi_{l}^{(+)}(k,r)\phi_{l'}^{(+)}(k,r)-j_{l}(kr)j_{l'}(kr)]dr,\label{gllp}
\end{eqnarray}
where $D_{l,l'}^{(m)}$ is defined as
\begin{equation}
D_{l,l'}^{(m)}=2i^{(l-l')}\int d\hat{{\bf r}}(3\cos^{2}\theta_{{\bf r}}-1)Y_{l'}^{m}(\hat{{\bf r}})^{*}Y_{l}^{m}(\hat{{\bf r}}).\label{dllm}
\end{equation}
Here $j_{l}(z)$ is the $l$-th order regular spherical Bessel function,
$\phi_{l}^{(+)}(k,r)$ is the short-range radial scattering wave function
in the $l$-th partial-wave channel, and satisfies $\langle{\bf r}|s_{{\bf k}}^{(+)}\rangle=\sqrt{\frac{2}{\pi}}\sum_{l,m}i^{l}\phi_{l}^{(+)}(|{\bf {\bf k}}|,r)Y_{l}^{m}(\hat{{\bf r}})Y_{l}^{m}(\hat{{\bf k}})^{*},$
with $|{\bf r}\rangle$ the eigen-state of inter-dipole relative coordinate
operator with eigen-value ${\bf r}$. In the derivation of Eq. (\ref{gllp}),
we have used the fact $\langle{\bf r}|s_{{\bf k}}^{(-)}\rangle=\langle{\bf r}|s_{-{\bf k}}^{(+)}\rangle^{*}$,
which is due to the boundary conditions satisfied by $|s_{{\bf k}}^{(\pm)}\rangle$.

Equation (\ref{gllp}) shows that the magnitude of $G_{l,l'}^{(m)}(k)$
is determined by the difference between the short-range scattering
wave function $\phi_{l(l')}^{(+)}(k,r)$ and the free wave function
$j_{l(l')}(kr)$ in the $l$-th ($l'$-th) partial wave channels.
In the following two sections, for simplicity, we assume that the SRI effects
in high-partial-wave channels  are very weak. Accordingly,
in the following two sections
we only consider the effects
induced by the difference between $\phi_{0}(k,r)$ and $j_{0}(kr)$
for dipolar bosons, and the effects from the difference between $\phi_{1}(k,r)$
and $j_{1}(kr)$ for dipolar fermions. In Sec. V we will consider a system where
the difference between $\phi_{2}^{(+)}(k,r)$ and $j_{2}(kr)$ cannot
be neglected.

\section{SRI-induced effects for bosonic dipoles}

In this section we consider an ultracold gas of polarized bosonic dipoles.
As shown above, we only consider the effect given by the difference
between $\phi_{0}(k,r)$ and $j_{0}(kr)$. Therefore,
terms $G_{l,l'}^{(m)}(k)$ are non-zero only for $(l,l')=(0,2)$ or
$(2,0)$. It is also easy to prove that $G_{0,2}^{(0)}(k)=G_{2,0}^{(0)}(k)$.
Thus, the SRI-induced effect for inter-dipole scattering is completely
described by the amplitude $G_{0,2}^{(0)}(k)$.

With a straightforward calculation based on Eq. (\ref{gllp}) (appendix A) , we
obtain 
\begin{equation}
G_{0,2}^{(0)}(k)=\frac{a_{d}}{3\sqrt{5}}\frac{ka}{(ika+1)},\label{g02}
\end{equation}
where $a$ is scattering length of the SRI. Furthermore, the amplitude
$P_{0,2}^{(0)}$ contributed by the FBA is
$P_{0,2}^{(0)}=-a_{d}/(3\sqrt{5})$. Therefore, $G_{0,2}^{(0)}(k)$
is comparable to $P_{0,2}^{(0)}$ only when $ka\rightarrow\infty$.
In this limit the SRI-induced effect is significant. On the other hand, when 
the $s$-wave scattering length $a$ is comparable to
or smaller than the dipolar length $a_{d}$, and the inter-atomic
momentum $k$ is much smaller than $1/a_{d}$, we have $|ka|<<1$. According to Eq. (\ref{g02}),
in this case
$G_{0,2}^{(0)}(k)$ is much smaller than $P_{0,2}^{(0)}$, and can be 
safely neglected. In these systems
the FBA is a very good approximation.
This conclusion is also supported by Fig. 1, where the behaviors of
$P_{0,2}^{(0)}$ and $G_{0,2}^{(0)}(k)$ for systems with different
scattering lengths are illustrated.

\begin{figure}[tbp]
\includegraphics[width=12cm]{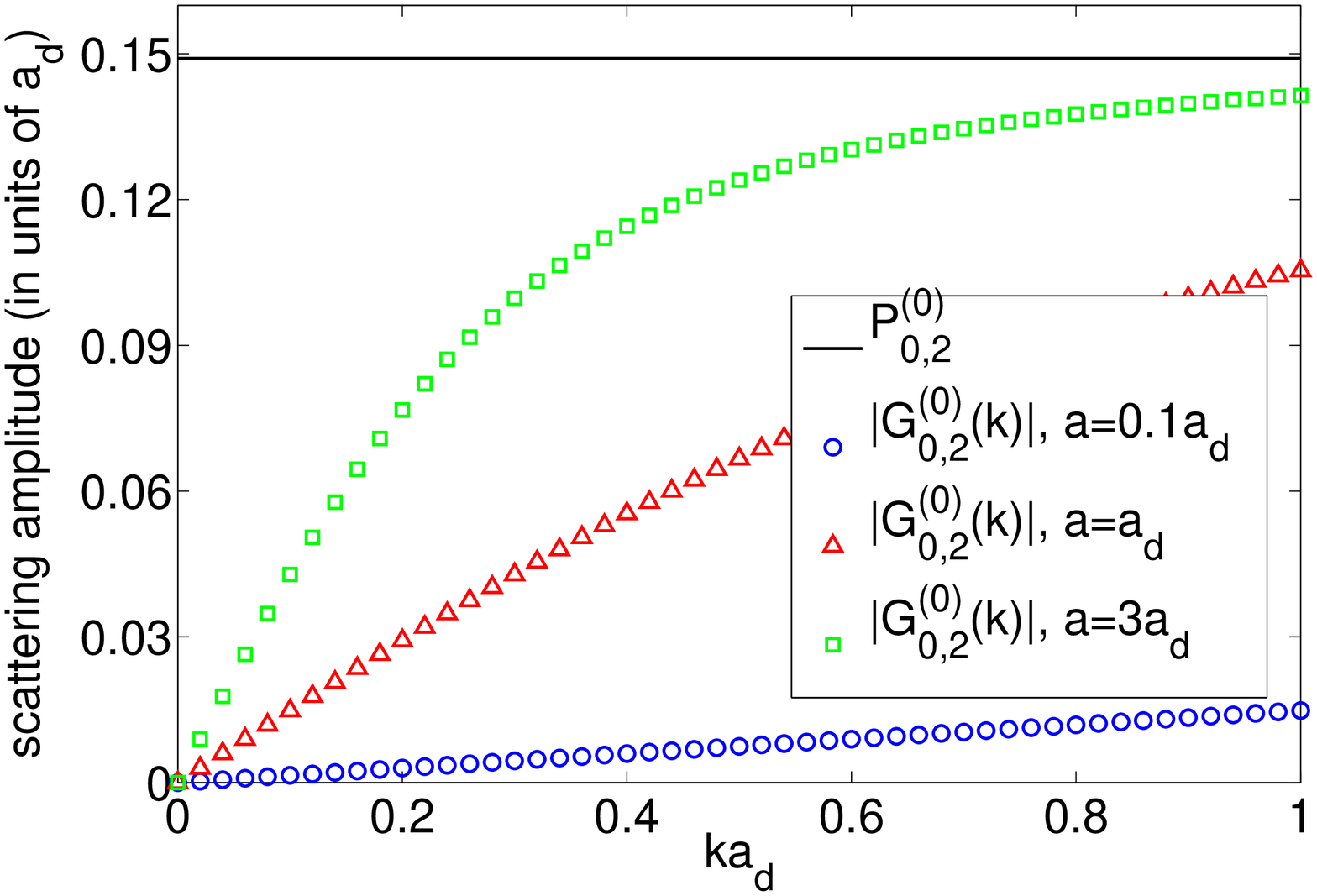}
\caption{(color online) The amplitudes $P_{0,2}^{(0)}$ (black solid line) and
$G_{0,2}^{(0)}(k)$ in ultracold gas of
bosonic dipoles with short-range scattering length $a=0.1a_{d}$ (blue
circles), $a=a_{d}$ (red triangles) and $a=3a_{d}$ (green squares). }
\end{figure}

\section{SRI-induced effects for fermionic dipoles}

\subsection{``Fast-decay'' type SRI}

Now we investigate the SRI-induced effects for the scattering of two identical
fermionic dipoles. We first assume that the SRI potential $V_{\rm sr}(r)$
is a ``fast-decay'' type potential, which decreases faster than
every power of $1/r$ as $r\rightarrow\infty$ and becomes negligible
when $r$ is larger than a characteristic distance $b$. In our system
the SRI-induced term $G_{l,l'}^{(m)}(k)$ can be obtained with an numerical
calculation based on Eq. (\ref{gllp}). Furthermore, in the low-energy
limit where the short-range $p$-wave scattering amplitude can be
approximated as $F_{1}^{\rm (sr)}(k)=-vk^{2}$, with $v$ the scattering
volume of the SRI \cite{lel}, we can obtain an approximate analytical expression
for $G_{l,l'}^{(m)}(k)$ (appendix B) with the help of the effective-range theory
\cite{taylor}: 
\begin{eqnarray}
G_{1,1}^{(m)}(k) & \approx & a_{d}\frac{v^{2}k^{2}}{r_{a}^{4}}\lambda_{m},\label{gdp}\\
G_{l,l'}^{(m)}(k) & \approx & 0,\ {\rm for}\ (l,l')\neq(1,1).\label{gdp2}
\end{eqnarray}
Here $\lambda_{0}=2/5$ and $\lambda_{\pm1}=-1/5$. $r_{a}$ is a
$k$-independent parameter determined by the short-range detail of
$V_{\rm sr}(r)$. It is defined as 
\begin{eqnarray}
r_{a}=\left[b^{-4}+\int_{0}^{b}\frac{\beta\left(r\right)^{2}}{r}dr\right]^{-1/4},
\end{eqnarray}
where $\beta(r)$ is the $k$-independent real function which satisfies
$\phi_{1}(k,r)\propto F_{1}^{\rm (sr)}(k)\beta(r)/k$ for $r\ll1/k$ (appendix
B). Equation (\ref{gdp}) yields that the generalized partial-wave scattering
amplitude modified by the SRI-induced amplitude $G_{l,l'}^{(m)}(k)$
is 
\begin{equation}
F_{1,1}^{(m)}(k)\approx-vk^{2}+a_{d}\left(1+\frac{v^{2}k^{2}}{r_{a}^{4}}\right)\lambda_{m}.\label{f11}
\end{equation}
Here we have also used the fact that $P_{1,1}^{(m)}=a_{d}\lambda_{m}$.

\begin{figure*}
\includegraphics[width=22cm]{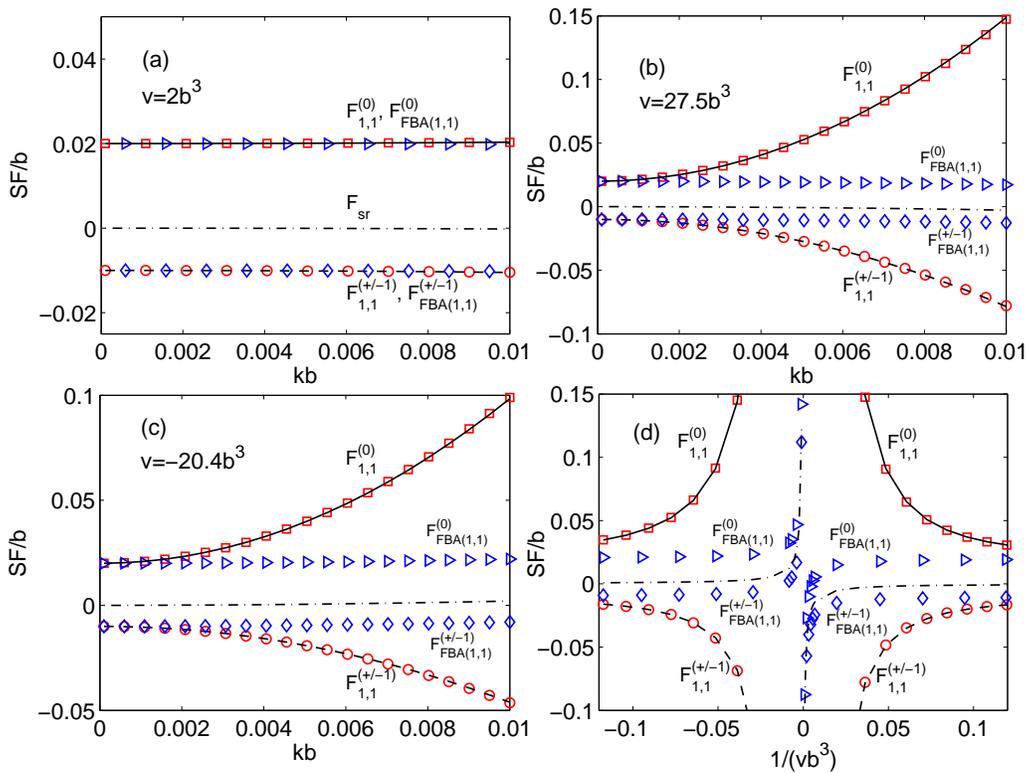} 
\caption{(color online) Scattering amplitudes (SFs) for ultracold fermions with square-well-type
SRI. Here we plot the DWBA scattering amplitude $F_{1,1}^{(m)}$ given by Eqs. (\ref{fp-1}, \ref{gllp}) (red squares for $m=0$ and
red circles for $m=\pm1$), $F_{1,1}^{(m)}$ given by Eq. (\ref{f11}) (black solid line for $m=0$ and black
dashed line for $m=\pm1$), the FBA scattering amplitude $F_{{\rm FBA}(1,1)}^{(m)}$ (blue triangles
for $m=0$ and blue diamonds for $m=\pm1$), and the short-range scattering amplitude $F_{1}^{\rm (sr)}$ (black
dashed-dotted line). The scattering amplitudes are plotted as functions of incident momentum
$k$ for the cases with scattering volume $v=2b^{3}$ (a), $v=27.5b^{3}$
(b) and $v=-20.4b^{3}$ (c), and as functions of scattering volume
for the case with $k=0.01/b$ (d). Here we only plot real part of
the scattering amplitudes, because in the parameter regions of these figures the value
of the imaginary parts of the scattering amplitudes
are at least $3$ orders of magnitude smaller than the real parts,
and thus can be neglected.}
\end{figure*}

Equations (\ref{gdp}) and (\ref{f11}) show that, when the scattering energy
is exactly zero, i.e., $k=0$, we have $G_{1,1}^{(m)}(k=0)=0$ and
thus the total scattering amplitude $F_{1,1}^{(m)}$ equals to $F_{{\rm FBA}(1,1)}^{(m)}$
from the FBA. However, for the cases with
finite incident momentum $k$, $F_{1,1}^{(m)}(k)$ is modified by the
SRI-induced scattering amplitude $G_{1,1}^{(m)}(k)$. In particular,
when $|v|\gg r_{a}^{4}/a_{d}$, $G_{1,1}^{(m)}(k)$ is much larger
than the short-range scattering amplitude $-vk^{2}$, and dominates
the variation of $F_{1,1}^{(m)}(k)$ with $k$. In that case the increasing
of $F_{1,1}^{(m)}(k)$ with $k$ is significantly enhanced by such SRI-induced
term.

This SRI-induced effect is important in systems where $|v|$ is
much larger than both $r_{a}^{4}/b$ and $r_{a}^{4}/a_{d}$. In these
systems, when $k$ is of the same order of magnitude as $r_{a}^{2}/|v|$
\cite{significant}, we have $|v|k^{2}\sim r_{a}^{4}/|v|\ll a_{d}$ and
$v^{2}k^{2}/r_{a}^{4}\sim1$. These facts yield $|G_{1,1}^{(m)}(k)|\sim |P_{1,1}^{(m)}|$ 
and $|F_{1}^{\rm (sr)}(k)|<<|P_{1,1}^{(m)}|$.
Namely, the SRI-induced amplitude $G_{1,1}^{(m)}(k)$ 
is
comparable to the term $P_{1,1}^{(m)}$ given by the FBA, while 
the short-range
scattering amplitude $F_{1}^{\rm (sr)}(k)$ is much smaller than $P_{1,1}^{(m)}$.
As a result, the FBA is no longer applicable,
although
the short-range
scattering amplitude is negligibly small.

In Fig. 2 we illustrate the above SRI-induced effect with a simple square-well
model of $V_{\rm sr}$, i.e., 
\begin{equation}
V_{\rm sr}(r)=
\begin{cases}
-V_0& \text{for}\ r<b;\\
0& \text{for}\ r>b.
\end{cases}
\end{equation}
In this model the scattering volume $v$ is determined by the depth
$V_{0}$ of the square well. We calculate the DWBA scattering amplitude
$F_{1,1}^{(m)}$ with both Eqs.
(\ref{fp-1}, \ref{gllp}), and Eq. (\ref{f11}).
For comparison, we also calculate the scattering amplitude $F_{{\rm FBA}(1,1)}^{(m)}$
from the FBA, and the short-range scattering
amplitude $F_{1}^{\rm (sr)}$. In our calculations we take $a_{d}=0.05b$.
As shown in Fig. 2, the FBA 
works very well when the scattering volume $v$ is small. Nevertheless, when $v$ is large,
$F_{1,1}^{(m)}$ significantly differs from $F_{{\rm FBA}(1,1)}^{(m)}$, although the short-range
scattering amplitude $F_{1}^{\rm (sr)}$ is still negligible. Furthermore,
it is also shown that our analytical expression (\ref{f11}) is always
a good approximation for $F_{1,1}^{(m)}$. For this system we also
calculate the amplitudes $G_{1,3}^{(m)}(k)$ and $G_{3,1}^{(m)}(k)$.
The calculation shows that, as predicted in Eq. (\ref{gdp2}), when
$F_{1}^{\rm (sr)}\lesssim a_{d}$ these amplitudes are 3 orders of magnitude
smaller than $P_{1,3}^{(m)}$ and $P_{3,1}^{(m)}$
from the FBA, and thus can be safely neglected.

\begin{figure}
\includegraphics[width=22cm]{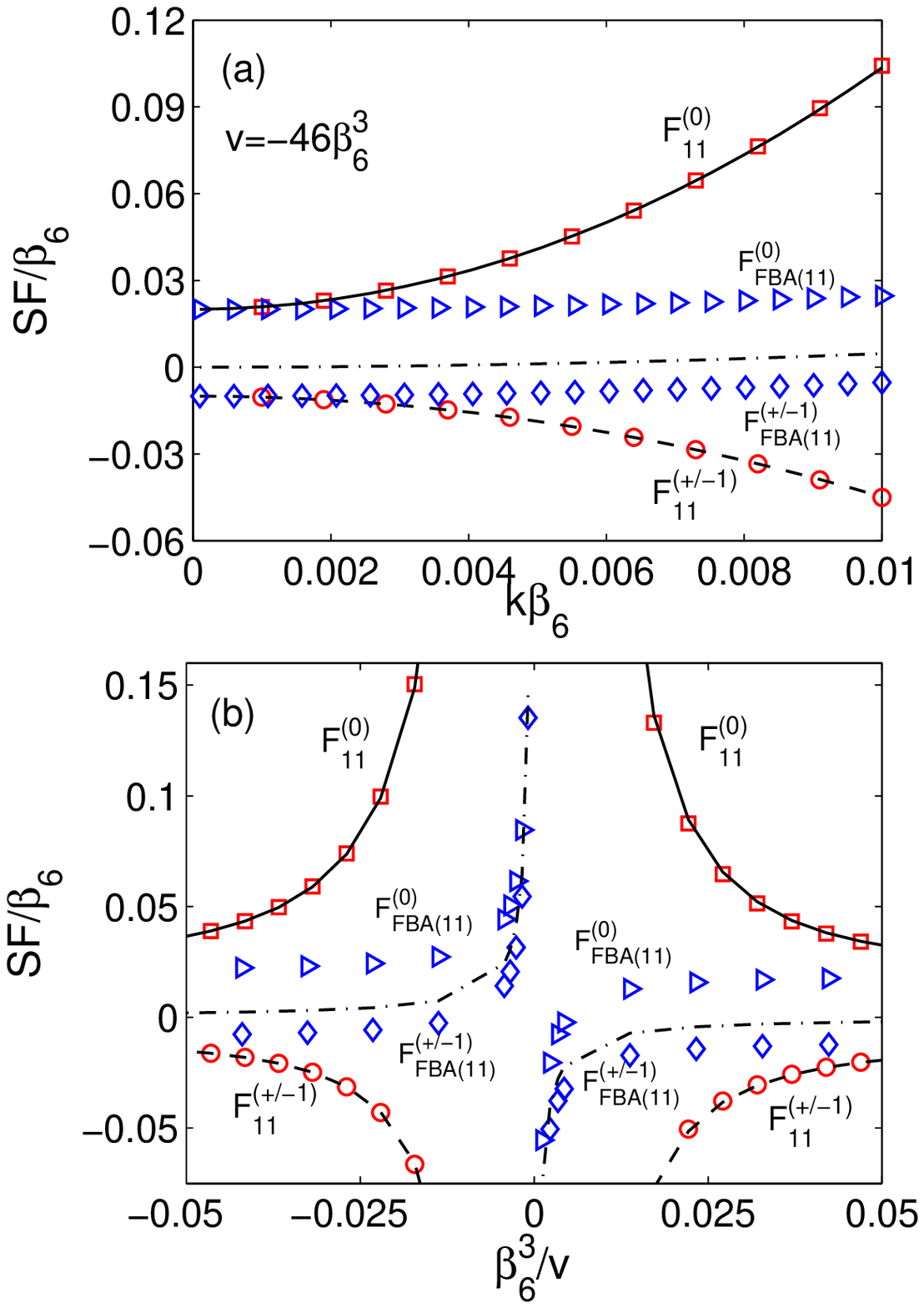}
\caption{(color online) Scattering amplitudes (SFs) for systems with van der
Waals type SRI. Here we plot scattering amplitudes as functions of $k$ for the case
with scattering volume $v=-46\beta_{6}^{3}$ (a), and as functions
of $1/v$ for the case with $k=0.01\beta_{6}$ (b). The black solid
line and dashed line are $F_{1,1}^{(m)}$ from Eq. (\ref{f11}) with
$r_{a}=0.4845\beta_{6}$. All other symbols have the same definitions
as in Fig. 2. Here we also neglect the imaginary parts of the scattering
amplitudes, which are at least 3 orders of magnitude smaller than the
real parts.}
\end{figure}

\begin{figure}
\includegraphics[width=22cm]{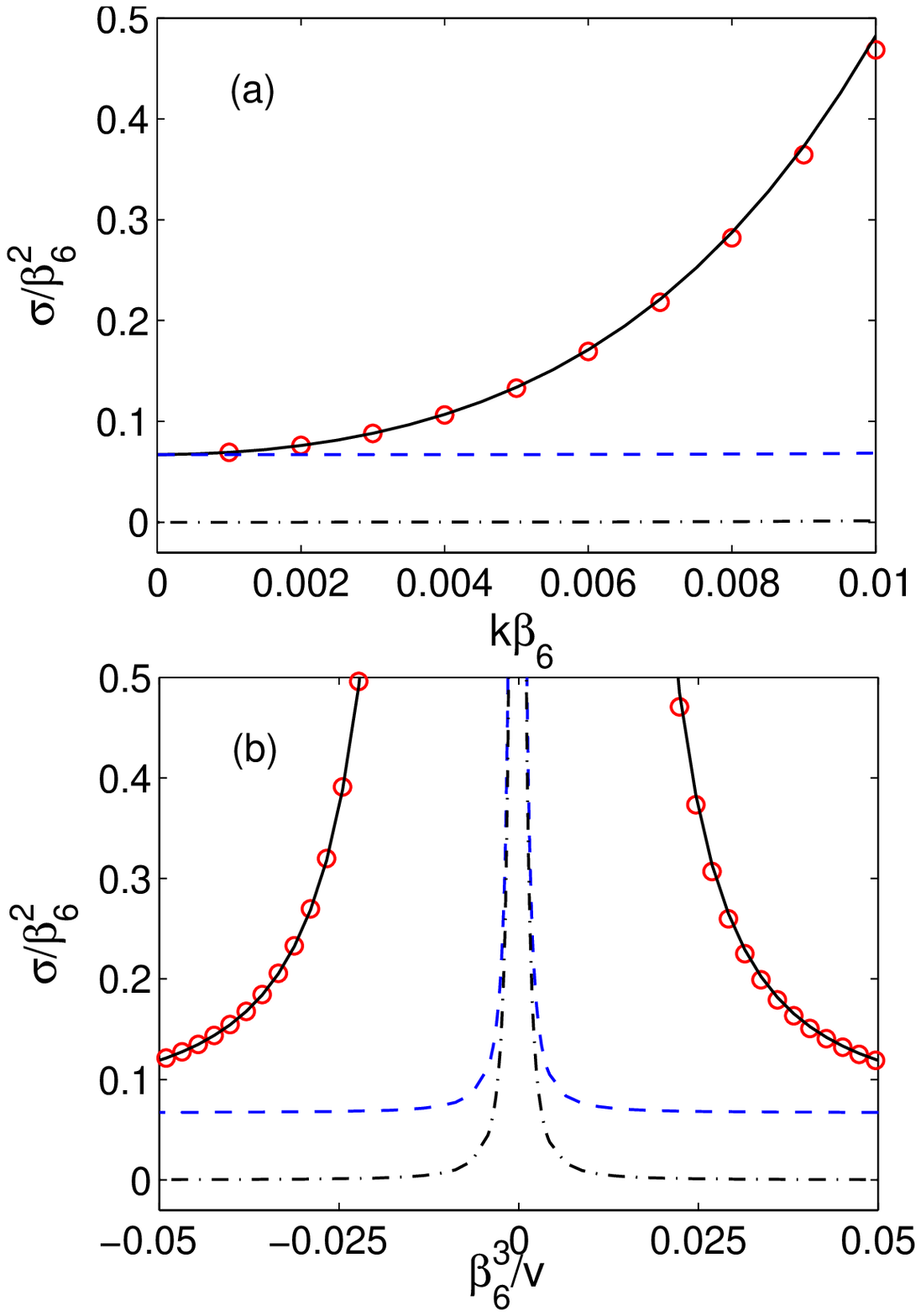}
\caption{(color online) Total scattering cross section of two fermionic dipoles
with van der Waals tpye SRI. Here we illustrate the cross section from
the numerical calculation based on the total scattering amplitude
$f$ in Eq. (\ref{fp}), with $F_{1,1}^{(m)}$ given by Eqs. (\ref{fp-1}) and (\ref{gllp}) (red circles), 
and Eq. (\ref{f11}) with $r_{a}=0.4845\beta_{6}$
(black solid line), as well as and the cross section given by $F_{{\rm FBA}(1,1)}^{(m)}$ (blue dashed line).
We also plot the scattering cross section for the SRI itself (black dashed-dotted
line). The cross sections are plotted as functions of incident momentum
$k$ for the case with $v=44.6\beta_{6}^{3}$ (a), and as functions
of $1/v$ for the case with $k=0.01\beta_{6}$ (b). }
\end{figure}

Here we also point out that Eq. (\ref{f11}) can be rewritten as
\begin{equation}
F_{1,1}^{(m)}(k)\approx-v_{{\rm eff}}^{(m)}k^{2}+P_{l,l'}^{(m)},\label{eq:-15}
\end{equation}
where 
\begin{equation}
v_{{\rm eff}}^{(m)}=v-\frac{a_{d}\lambda_{m}}{r_{a}^{4}}v^{2}.\label{f11-1}
\end{equation}
With this result, the effect from the amplitude $G_{1,1}^{(m)}(k)$
can also be understood as ``DDI-induced modification of the scattering
volume''. Namely, due to the DDI-SRI interplay, the effective scattering
volume of the two dipoles is shifted from $v$ to $v_{{\rm eff}}^{(m)}$,
which depends on the quantum number $m$ for the relative angular
momentum along the $z$-direction. In previous works people have studied
the variation of scattering volume by DDI in the \emph{close} channel
of a $p$-wave Feshbach resonance \cite{bohnandjin}. Since our calculation
is based on a single-channel model, the shift of the scattering volume
we study here is due to the DDI in the \emph{open} channel.

\subsection{\emph{``}van der Waals type'' SRI}

Now we consider the case where the SRI is a ``van der Waals type
potential'', which has the behavior $-\beta_{6}^{4}/r^{6}$ when
$r$ is larger than a distance $c$. Here we assume both $\beta_{6}$
and $c$ are much smaller than $1/k$. In this case, the effective-range
theory cannot be used to study the scattering problem for $V_{\rm sr}$
\cite{Bo Gao}. As a result, we cannot use analytical techniques to
derive approximate expressions for the scattering amplitudes. Thus,
we study the SRI-induced effect via direct numerical calculation for
$F_{1,1}^{(m)}(k)$ with Eq. (\ref{fp-1}). In our calculation we
consider the limit $c\rightarrow0$. In such a limit the scattering-state
wave function $\langle{\bf r}|s_{{\bf k}}^{(\pm)}\rangle$ is the
superposition of the incident plane wave and the exact solutions of
the $p$-wave Schr$\ddot{\rm o}$dinger equation with van der Waals potential,
which is analytically obtained by Bo Gao in Ref. \cite{Bo Gao 2nd}. The superposition coefficient is determined by the scattering volume
of the system \cite{Bo Gao}.

In Fig. 3 we show our numerical results for the scattering amplitude
$F_{1,1}^{(m)}(k)$ given by the DWBA and the $F_{{\rm FBA}(1,1)}^{(m)}(k)$
given by the FBA, and the short-range scattering
amplitude $F_{1}^{\rm (sr)}$. In our calculation we take $a_{d}=0.05\beta_{6}$.
It is shown that, in our system the amplitude $G_{1,1}^{(m)}(k)$ can also be significantly large
when the $p$-wave scattering volume is large enough. In these
cases $F_{1,1}^{(m)}(k)$ cannot be approximated as $F_{{\rm FBA}(1,1)}^{(m)}(k)$,
even if $F_{1}^{\rm (sr)}$ is much smaller than $a_{d}$.

Furthermore, we are surprised to find that our numerical result also agrees
very well with expression (\ref{f11}) with $r_{a}=0.4845\beta_{6}$
(Fig. 3). This fact implies that, although Eq. (\ref{f11}) is derived
for the system with fast-decay type SRI, it can still be used to describe
the scattering amplitude for the system with van der Waals type SRI.
Nevertheless, in the current system $r_{a}$ becomes a $k$-independent
fitting parameter.

The SRI-induced effect we discussed above can be experimentally
observed via the measurement of scattering cross sections of ultracold
fermionic dipoles. In Fig. 4 we illustrate the total cross section
$\sigma\equiv\int d\hat{{\bf k}}_{i}d\hat{{\bf k}}_{f}|f_{F}({\bf k}_{i},{\bf k}_{f})|^{2}/4\pi$, 
where $f_{F}({\bf k}_{i},{\bf k}_{f})=[f({\bf k}_{i},{\bf k}_{f})-f({\bf k}_{i},-{\bf k}_{f})]/\sqrt{2}$,
for ultracold fermonic dipoles with van der Waals type SRI with $a_{d}=0.05\beta_{6}$.
It is shown that, when the scattering volume of the SRI is large enough,
the scattering cross section is significantly enhanced by the SRI-induced
effect.

\section{SRI-induced effects for bosonic dipoles with a strong d-wave SRI}

In above two sections, we have neglected the difference between the short-range
scattering wave function $\phi_{l}^{(+)}(k,r)$ and the free wave
function $j_{l}(kr)$ in high partial-wave channels. Such approximation
is applicable when the low-energy short-range scattering effects in
these channels are very weak. In this section we consider the ultracold
bosonic dipoles with a ``fast-decay''-type SRI which has strong effect in the
$d$-wave channel. For such a system, both the difference betwen $\phi_{0}^{(+)}(k,r)$
and $j_{0}(kr)$ and the difference between $\phi_{2}^{(+)}(k,r)$
and $j_{2}(kr)$ should be taken into account. We find that, under
particular conditions, the generalized partial-wave scattering amplitudes
$F_{2,2}^{(m)}(k)$ and $F_{2,0}^{(m)}(k)$ can differ from the result
given by FBA even when $k=0$.

According to the effective range theory, for a ``fast-decay'' type
SRI the low-energy $d$-wave scattering amplitude $F_{2}^{\rm (sr)}(k)$
is given by
\begin{equation}
F_{2}^{\rm (sr)}(k)=\frac{-1}{ik+\frac{1}{wk^{4}}+\frac{1}{s_{*}k^{2}}+\frac{1}{u}},\label{eq:-16}
\end{equation}
with parameters $w$, $s_{*}$ and $u$ determined by the details of
the SRI. In this section we consider the systems where $s_{*}$ and
$u$ have the same order of magnitude as $b$. For these systems,
with direct calculations (appendix C) we find that the SRI-induced
amplitudes can be approximately expressed as
\begin{eqnarray}
G_{2,2}^{(m)}(k) & \approx & a_{d}\left[\frac{3}{2}\frac{F_{2}^{\rm (sr)}(k)^{2}}{r_{b}^{6}k^{4}}+\frac{2}{5}\frac{F_{2}^{\rm (sr)}(k)}{b}\right]D_{2,2}^{(m)};\label{g22}\\
G_{2,0}^{(m)}(k) & \approx & a_{d}\left[\frac{F_{2}^{\rm (sr)}(k)}{r_{c}(k)^{3}k^{2}}+\frac{F_{0}^{\rm (sr)}(k)F_{2}^{\rm (sr)}(k)}{kb^{3}}\right],
 \label{g20}
\end{eqnarray}
$G_{0,2}^{(m)}(k)=G_{2,0}^{(m)}(k)$, and $G_{l,l'}^{(m)}(k)\approx0$
for $(l,l')\neq(2,2)$, $(2,0)$ and $(0,2)$. Here the parameters
$D_{2,2}^{(m)}$ and $D_{2,0}^{(m)}$ are defined in Eq. (\ref{dllm}).
The ranges $r_{b}$ and $r_{c}(k)$ are defined as $r_{b}=(b^{-6}+\frac{2}{3}\int_{0}^{b}\beta_{2}(r)^{2}/rdr)^{-1/6}$
and $r_{c}(k)=(b^{-3}+\int_{0}^{b}\phi_{0}^{(+)}(k,r)\beta_{2}(r)/rdr)^{-1/3}$,
respectively, with function $\beta_{2}(r)$ satisfying $\phi_{2}^{(+)}(k,r\lesssim b)=F_{2}^{\rm (sr)}(k)\beta_{2}(r)/k^{2}.$ 

\begin{figure*}
\includegraphics[width=22cm]{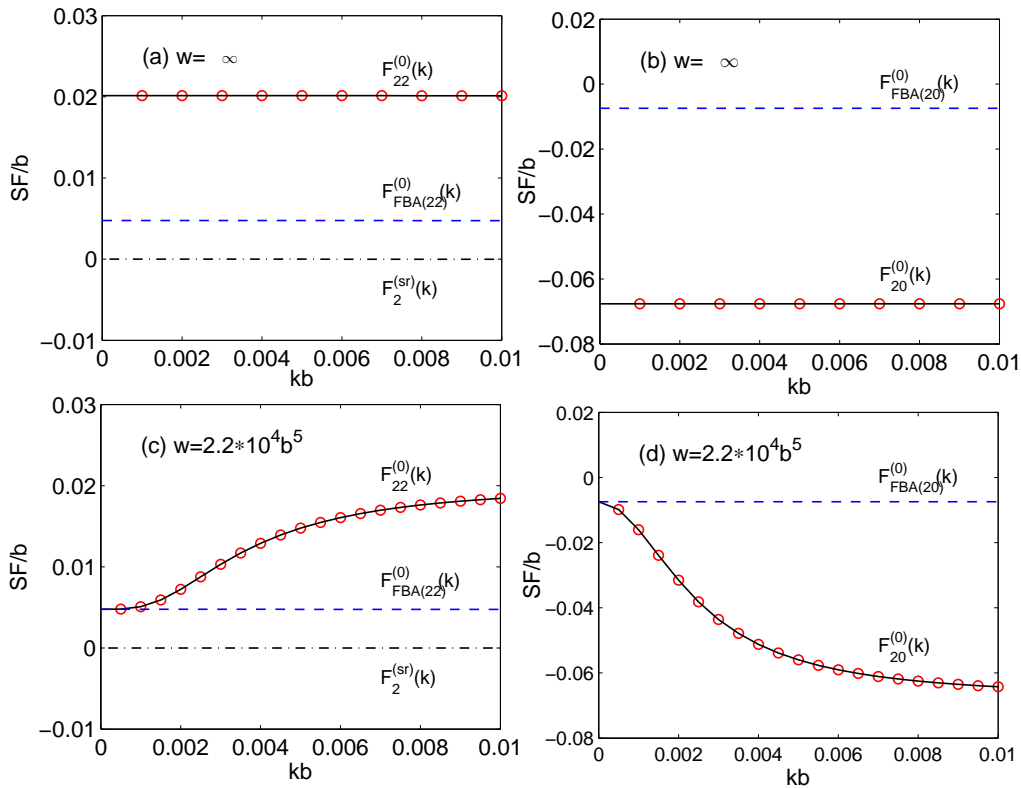} 
\caption{(color online) Scattering amplitudes (SFs) for bosons with square-well-type
SRI. Here we consider the cases with $w=\infty$ (a, b) and $w=2.2\times10^{4}b^{5}$
(c, d). For each case we plot $F_{2,2}^{(0)}$ and $F_{2,0}^{(0)}$
obtained from the numerical calculations based on Eqs. (\ref{fp-1}) and (\ref{gllp})
(red circles), and the calculation based on the approximated expressions (\ref{g22}) and (\ref{g20})
(black solid line below the red circles), as well as the results from
the FBA (blue dashed lines). We also illustrate the $d$-wave
short-range scattering amplitude $F_{2}^{\rm (sr)}$ (black dashed-dotted
line). Here we also neglect the imaginary parts of the scattering
amplitudes, which are at least 6 orders of magnitude smaller than the
real parts.}
\end{figure*}

Using Eq. (\ref{eq:-16}) and Eqs. (\ref{g22}) and (\ref{g20}), we find
that when $w=\infty$ the SRI-induced amplitudes $G_{2,2}^{(m)}(k)$
and $G_{2,0}^{(m)}(k)$ are non-zero in the zero-energy limit $k=0$,
i.e., we have 
\begin{eqnarray}
G_{2,2}^{(m)}(k=0) & \approx& \frac{3a_{d}}{2r_{b}^{6}}s_{*}^{2}D_{2,2}^{(m)};\label{eq:-17}\\
G_{2,0}^{(m)}(k=0) & \approx& \frac{a_{d}s_{*}}{r_{c}(0)^{3}}D_{2,0}^{(m)}.\label{eq:-18}
\end{eqnarray}
Therefore, when $w=\infty$, the SRI-induced amplitudes 
$G_{2,2}^{(m)}$ and $G_{2,0}^{(m)}$
 can significantly modify the generalized
parital-wave scattering amplitudes $F_{2,2}^{(m)}$ and $F_{2,2}^{(m)}$
in the zero-energy limit, although in this limit the $d$-wave scattering
amplitude $F_{2}^{\rm (sr)}$ of the SRI is exactly zero. It is pointed out that,
this effect does not occur in the systems we discussed in the above two sections.
According to Eqs. (\ref{g02}) and (\ref{gdp}), in these systems, 
when the short-range scattering amplitudes are exactly zero (i.e., $a=0$ or $k=0$),
the amplitudes $G_{0,2}^{(m)}$, $G_{2,0}^{(m)}$ and  $G_{1,1}^{(m)}$ are also zero. As a result,
the DWBA and the FBA give the same scattering amplitudes.

When $w$ is finite, we have $G_{2,2}^{(m)}(k=0)=G_{2,0}^{(m)}(k=0)=0$.
Thus, when $k=0$, the scattering amplitude given by the DWBA equals to the one
from the FBA. Nevertheless, when $k$ is finite and $w$
is much larger than $b^{5}$, the SRI-induced amplitudes $G_{2,2}^{(m)}(k)$
and $G_{2,0}^{(m)}(k)$ can still be significantly large even if $F_{2}^{\rm (sr)}(k)$
is much smaller than the dipolar length $a_{d}$. This effect is quite similar to the one we discussed in the above section.

In Fig. 5 we illustrate the above SRI-induced effects with the square-well
model of $V_{\rm sr}$. It is clearly shown that when $w=\infty$, the
total scattering amplitudes differ from the results given by the FBA in both
the cases with $k=0$ and $k\neq0$. When $w$
takes a large finite value, the FBA works well when
$k=0$, while the SRI-induced effects can be significant when $k$ is
finite.

\section{conclusions and discussions}

In this work we investigate the interplay of SRI and DDI in the scattering
of two polarized weak dipoles. We show that this interplay can be
safely neglected for ultracold bosonic dipoles when the scattering
length of the SRI is much smaller than the dipolar length $a_{d}$,
and the effects of SRI in high partial-wave channels are negligible.
Nevertheless, for ultracold fermionic dipoles with large scattering
volume, or ultracold bosonic dipoles with a strong ``fast-decay''
type SRI in the $d$-wave channel, the SRI-DDI interplay is important even when
the scattering amplitude for the SRI is much smaller than $a_{d}$.
In these systems the FBA is not applicable.
We find analytical expressions for the scattering amplitudes of these
systems. With these expressions one can construct appropriate effective
interaction potentials in the many-body effective Hamiltonians.

\begin{acknowledgments}
This work has been supported by National Natural Science Foundation
of China under Grants No. 11222430, NKBRSF of China under Grants No.
2012CB922104. Peng Zhang also thank Hui Zhai, Shizhong Zhang, Wenxian
Zhang, Doerte Blume, Shina Tan, Li Ge, Su Yi, Stefano Chesi, Ran Qi, Zhenhua Yu,
Li You, Paul. S. Julienne, Wei Zhang, Hui Hu, Gang Chen and Hossein Sadeghpour
for helpful discussions. 
\end{acknowledgments}
\appendix
\addcontentsline{toc}{section}{Appendices}\markboth{APPENDICES}{}
\begin{subappendices}

\section{SRI-induced amplitudes for bosonic dipoles}

In this appendix we prove Eq. (\ref{g02}) in Sec. III of our main text. According
to Eq. (\ref{gllp}), the amplitude $G_{0,2}^{(0)}(k)$ is given by
$G_{0,2}^{(0)}(k)=I_{1}+I_{2}$, where 
\begin{eqnarray}
I_{1} & = & -\frac{4a_{d}}{\sqrt{5}}\int_{0}^{b}\frac{1}{r}[\phi_{0}^{(+)}(k,r)\phi_{2}^{(+)}(k,r)-j_{0}(kr)j_{2}(kr)]dr;\nonumber \\
\\
I_{2} & = & -\frac{4a_{d}}{\sqrt{5}}\int_{b}^{\infty}\frac{1}{r}[\phi_{0}^{(+)}(k,r)\phi_{2}^{(+)}(k,r)-j_{0}(kr)j_{2}(kr)]dr,\nonumber \\
\end{eqnarray}
with $b$ the range of the SRI. Here we have used the fact $D_{0,2}^{(0)}=-4/\sqrt{5}$.

As shown in Sec. II of our main text, we neglect the difference between
$\phi_{l}^{(+)}(k,r)$ and $j_{l}(kr)$ with $l>0$.
Therefore, in above equations we can replace $\phi_{2}^{(+)}(k,r)$
with $j_{2}(kr)$. Furthermore, in the realistic systems of ultracold
dipolar gases, the integration $I_{1}$ for the wave functions in the
region $r<b$ can be neglected. Thus, the function $G_{0,2}^{(0)}(k)$
can be approximated as $G_{0,2}^{(0)}(k)\approx I_{2}$. In addition,
in the region $r>b$ the $s$-wave scattering wave function $\phi_{0}^{(+)}(k,r)$
can be expressed as $\phi_{0}^{(+)}(k,r>b)=j_{0}(kr)+kF_{0}(k)e^{ikr}$,
with $F_{0}(k)$ the $s$-wave scattering amplitude for the SRI. Therefore,
we have 
\begin{equation}
G_{0,2}^{(0)}(k)\approx I_{2}=-\frac{2d^{2}}{\sqrt{5}}\int_{b}^{\infty}\frac{1}{r}kF_{0}(k)e^{ikr}j_{2}(kr)dr.\label{eq:-14}
\end{equation}
For ultracold gases with $k\ll1/b$, we can approximate the lower
limit $b$ of the integration in above equation as zero. Under this
approximation and the relation $F_{0}(k)=-1/[ik+1/a]$, with $a$ the
scattering length, we get Eq. (\ref{g02}) in our main text.

\section{SRI-induced amplitudes for fermionic dipoles}

In this appendix we calculate the SRI-induced amplitude $G_{l,l'}^{(m)}(k)$
for fermionic dipoles with a ``fast-decay'' type SRI, and prove Eq.
(\ref{gdp}) in our main text. According to the effective-range theory
\cite{taylor}, we have 
\begin{equation}
F_{1}^{\rm (sr)}(k)=-\frac{1}{ik+\frac{1}{vk^{2}}+\frac{1}{r_{{\rm eff}}}},\label{f1sr}
\end{equation}
where $v$ is the scattering volume and $r_{{\rm eff}}$ is the effective
range. Here we assume $r_{{\rm eff}}$ is of the same order of magnitude
as $b$. In our calculation we consider the low-enegy limit where
$k$ is much smaller than both $1/b$ and $\sqrt{|r_{{\rm eff}}/v|}$.
In this limit the $p$-wave short-range scattering amplitude given
by Eq. (\ref{f1sr}) can be approximated as $F_{1}^{\rm (sr)}(k)=-vk^{2}$. 

As shown in Sec. II, for ultracold dipolar fermions we only consider
the amplitudes induced by the difference between $\phi_{1}(k,r)$
and $j_{1}(kr)$, i.e., the amplitudes $G_{1,1}^{(m)}(k)$, $G_{3,1}^{(m)}(k)$,
and $G_{1,3}^{(m)}(k)$. Since $V_{\rm sr}$ is negligible in the region
$r>b,$ in this region we have 
\begin{eqnarray}
\phi_{1}^{(s)}(k,r) & \equiv & \phi_{1}^{(+)}(k,r)-j_{1}(kr)\nonumber \\
 & = & -i\frac{F_{1}^{\rm (sr)}(k)}{r}\left(1+\frac{i}{kr}\right)e^{ikr}.\label{phip-2}
\end{eqnarray}
On the other hand, Eq. (\ref{gllp}) in our main text yields
\begin{equation}
G_{1,1}^{(m)}(k)=a_{d}D_{1,1}^{(m)}\left(\alpha_{1}+\alpha_{2}+\alpha_{3}\right),\label{a11}
\end{equation}
where 
\begin{eqnarray}
\alpha_{1} & = & \int_{b}^{\infty}\phi_{1}^{(s)}(k,r)^{2}\frac{1}{r}dr\nonumber \\
 &  & +2\int_{b}^{\infty}\phi_{1}^{(s)}(k,r)j_{1}\left(kr\right)\frac{1}{r}dr;\label{eq:}\\
\alpha_{2} & = & -\int_{0}^{b}j_{1}\left(kr\right)^{2}\frac{1}{r}dr;\label{eq:-1}\\
\alpha_{3} & = & \int_{0}^{b}\phi_{1}^{(+)}(k,r)^{2}\frac{1}{r}dr.\label{eq:-2}
\end{eqnarray}
Using Eq. (\ref{phip-2}), we directly obtain 
\begin{equation}
\alpha_{1}=8\pi a_{d}\left[\frac{F_{1}^{\rm (sr)}(k)^{2}}{4b^{4}k^{2}}+\frac{2}{3}\frac{F_{1}^{\rm (sr)}(k)}{b}\right].\label{eq:-3}
\end{equation}
In addition, since $kb\ll1$, $|\alpha_{2}|$ is much less than unit.
Therefore, in $G_{1,1}^{(m)}(k)$ the contribution given by $\alpha_{2}$
are much smaller than $P_{1,1}^{(m)}=a_{d}D_{1,1}^{(m)}/4$, and thus
can be neglected.

The factor $\alpha_{3}$ can be calculated as following. It is clear
that $\phi_{1}^{(+)}(k,r)$ satisfies the radial Schr$\ddot{\rm o}$dinger equation
\begin{equation}
\left(-\frac{1}{r}\frac{d^{2}}{dr^{2}}r+\frac{2}{r^{2}}+V_{\rm sr}\right)\phi_{1}^{(+)}(k,r)=k^{2}\phi_{1}^{(+)}(k,r)\label{se-1}
\end{equation}
with the boundary condition $\phi_{1}^{(+)}(k,0)=0$. When $r\ll1/k$,
one can neglect the term $k^{2}\phi_{1}^{(+)}(k,r)$ in this equation and thus obtain
$\phi_{1}^{(+)}(k,r)=\alpha(k)\beta(r)$, where $\beta(r)$ is a $k$-independent
real function. On the other hand, using the facts $b\sim r_{{\rm eff}}$
and $kb\ll1$ we find that, when
$r$ is both larger than $b$ and of the same order of magnitude as
$b$ (e.g., $r=2b$), we have
\begin{eqnarray}
\phi_{1}^{(+)}(k,r) & = & j_{1}(kr)-i\frac{F_{1}^{\rm (sr)}(k)}{r}\left(1+\frac{i}{kr}\right)e^{ikr}\nonumber \\
 & \approx & \frac{F_{1}^{\rm (sr)}(k)}{k}\left(\frac{1}{r^{2}}+\frac{k^{2}}{2}+\frac{ik^{3}r}{3}+\frac{k^{2}r}{3F_{1}^{\rm (sr)}(k)}\right).\nonumber \\
\label{eq:-10}
\end{eqnarray}
The fact $k\ll1/b$ yields $k^{2}/2\ll1/r^{2}$ and $k^{3}r/3\ll1/r^{2}$.
These results and $F_{1}^{\rm (sr)}(k)=-vk^{2}$ imply
\begin{equation}
\phi_{1}^{(+)}(k,r)\approx\frac{F_{1}^{\rm (sr)}(k)}{k}\left(\frac{1}{r^{2}}-\frac{r}{3v}\right).\label{eq:-12}
\end{equation}
Therefore, we have $\alpha(k)=F_{1}^{\rm (sr)}(k)/(2k)$, and thus 
\begin{equation}
\alpha_{3}=8\pi a_{d}\left(\frac{F_{1}^{\rm (sr)}(k)}{2k}\right)^{2}\int_{0}^{b}\frac{\beta\left(r\right)^{2}}{r}dr.\label{eq:-4}
\end{equation}
Using the above results for $\alpha_{1,2,3}$, we obtain 
\begin{equation}
G_{1,1}^{(m)}(k)=a_{d}D_{1,1}^{(m)}\left[\frac{v^{2}k^{2}}{4r_{a}^{4}}-\frac{2vk^{2}}{3b}\right],\label{biga11}
\end{equation}
where the length $r_{a}$ is defined as 
\begin{equation}
r_{a}=\left[\frac{1}{b^{4}}+\int_{0}^{b}\frac{\beta\left(r\right)^{2}}{r}dr\right]^{-1/4}.\label{eq:-5}
\end{equation}
It is clear that we have $r_{a}\lesssim b$. Here we have used the
fact that $F_{1}^{\rm (sr)}(k)=-vk^{2}$.

We can calculate $G_{(1,3)}^{(m)}(k)$ and $G_{(1,3)}^{(m)}(k)$ with
the similar approach as above. These calculations give 
\begin{equation}
G_{(1,3)}^{(m)}(k)=G_{(1,3)}^{(m)}(k)=a_{d}D_{1,3}^{(m)}(\beta_{1}+\beta_{2}),\label{biga13}
\end{equation}
where 
\begin{eqnarray}
\beta_{1} & = & \frac{i}{36}kF_{1}^{\rm (sr)}(k),\label{eq:-11}\\
\beta_{2} & = & \frac{1}{105}\left(F_{1}^{\rm (sr)}(k)k^{2}\right)\int_{0}^{b}\beta\left(r\right)r^{2}dr.\label{eq:-7}
\end{eqnarray}
Here we have neglected the terms which is much smaller than $P_{1,3}^{(m)}=P_{3,1}^{(m)}=a_{d}D_{1,1}^{(m)}/36$.

Furthermore, in the low-enegy limit $k\ll\sqrt{|r_{{\rm eff}}/v|}$
we have 
\begin{equation}
\left\vert \frac{vk^{2}}{b}\right\vert \ll\left\vert \frac{r_{{\rm eff}}}{b}\right\vert \sim1\label{eq:-6}
\end{equation}
and 
\begin{equation}
|kF_{1}^{\rm (sr)}(k)|\sim|vk^{3}|\ll|r_{{\rm eff}}k|\sim|kb|\ll1,\label{eq:-13}
\end{equation}
which yields 
\begin{eqnarray}
a_{d}D_{1,1}^{(m)}|\frac{2vk^{2}}{3b}| & \ll & P_{1,1}^{(m)},\ |a_{d}D_{1,3}^{(m)}\beta_{1}|\ll P_{1,3}^{(m)}.\nonumber \\
\label{aa}
\end{eqnarray}
In addition, in the practical cases, the short-range interaction potentials
$V_{\rm sr}$ between ultracold dipolar atoms or molecules are deep potential
wells. As a result, in the region $r<b$, the low-energy wave function
$\beta(r)$ is a rapid oscillating function with small amplitude.
Here we estimate the upper limit of $|\beta_{2}|$ as $|\frac{1}{105}F_{p}\left(k\right)\int_{0}^{b}r^{2}dr|\cdot|\beta\left(r\sim b\right)|$.
Using the facts $|\beta(r\sim b)|\lesssim|\frac{1}{b^{2}}|+|\frac{b}{3v}|$
and $kb\ll1$, we find that in the low-energy limit we have 
\begin{equation}
|a_{d}D_{1,3}^{(m)}\beta_{2}|\ll P_{1,3}^{(m)}.\label{bb}
\end{equation}

Using Eqs. (\ref{aa}, \ref{bb}), we find that in the low-energy
limit Eqs. (\ref{biga11}, \ref{biga13}) can be simplified as 
\begin{eqnarray}
G_{1,1}^{(m)}(k) & \approx & a_{d}\frac{v^{2}k^{2}}{r_{a}^{4}}\lambda_{m},\ G_{1,3}^{(m)}(k)=G_{3,1}^{(m)}(k)\approx0,\nonumber \\
\label{eq:-12-1}
\end{eqnarray}
with $\lambda_{m}=8/5$ for $m=0$ and $\lambda_{m}=-4/5$ for $m=\pm1$.
That is Eq. (\ref{gdp}) in Sec. IV.

\section{SRI-induced amplitudes for bosonic dipoles with a strong d-wave SRI}

In this appendix we prove Eqs. (\ref{g22}) and (\ref{g20}) in our
main text. According to the effective-range theory \cite{taylor},
the $d$-wave scattering wave function for the ``fast-decay'' type
SRI satisfies
\begin{equation}
\phi_{2}^{(+)}(k,r>b)=j_{2}(kr)+F_{2}^{\rm (sr)}(k)kh_{2}^{(+)}(k),\label{phi2}
\end{equation}
with the scattering amplitude $F_{2}^{\rm (sr)}(k)$ given by Eq. (\ref{eq:-16}).
Here we assume the parameters $s_{*}$ and $u$ have the same order
of magnitude as $b$. In addition, it is clear that $\phi_{2}^{(+)}(k,r)$
satisfies the radial Schr$\ddot{\rm o}$dinger equation 
\begin{equation}
\left(-\frac{1}{r}\frac{d^{2}}{dr^{2}}r+\frac{6}{r^{2}}+V_{\rm sr}\right)\phi_{2}^{(+)}(k,r)=k^{2}\phi_{2}^{(+)}(k,r),\label{se-1-1}
\end{equation}
with boundary condition $\phi_{2}^{(+)}(k,0)=0$. In the region $r\ll1/k$,
one can neglect the term $k^{2}\phi_{2}^{(+)}(k,r)$ and thus obtain
$\phi_{2}^{(+)}(k,r)=\alpha_{2}(k)\beta_{2}(r)$, where $\beta_{2}(r)$
is a $k$-independent real function. On the other hand, using the
facts that $b$ and $s_{*}$ and $u$ are on the same order of magnitude
and $kb\ll1$, we find that when $r$ is both larger than $b$ and
on the same order of magnitude as $b$ (e.g., $r=2b$), we have
\begin{equation}
\phi_{2}^{(+)}(k,r)\approx\frac{F_{2}^{\rm (sr)}(k)}{k^{2}}\left(\frac{3}{r^{3}}-\frac{r^{2}}{15w}\right).\label{eq:-19}
\end{equation}
Therefore, the funciton $\alpha_{2}(k)$ can be chosen as $\alpha_{2}(k)=F_{2}^{\rm (sr)}(k)/k^{2}$.
Namely, we have 
\begin{equation}
\phi_{2}^{(+)}(k,r\lesssim b)=\frac{F_{2}^{\rm (sr)}(k)}{k^{2}}\beta_{2}(r).\label{eq:-20}
\end{equation}
Substituting Eqs. (\ref{phi2}, \ref{eq:-20}) into Eq. (\ref{gllp}) in our main text,
we can straightforwardly prove Eq. (\ref{g22}) and (\ref{g20}) in our main text.
In addition, we can also obtain $G_{2,4}^{(m)}(k)=G_{4,2}^{(m)}(k)=ia_{d}kF_{2}^{\rm (sr)}(k)D_{2,4}^{(m)}/72$.
Since $(s_{*},u)\sim b$ and $kb<<1$, we have $F_{2}^{\rm (sr)}(k)<<b$
and thus $|kF_{2}^{\rm (sr)}(k)|<<1$. Therefore, the amplitudes $G_{2,4}^{(m)}(k)$
and $G_{4,2}^{(m)}(k)$ can be neglected.

\end{subappendices}


\begin{thebibliography}{10}
\bibitem{crbecexp1} A. Griesmaier, J. Werner, S. Hensler, J. Stuhler,
T. Pfau, Phys. Rev. Lett. \textbf{94}, 160401 (2005).

\bibitem{crbecexp2} J. Stuhler, A. Griesmaier, T. Koch, M. Fattori,
T. Pfau, Phys. Rev. Lett. \textbf{95}, 150406 (2005).

\bibitem{crbecexp3} A. Griesmaier, J. Stuhler, T. Koch, M. Fattori,
T. Pfau, S. Giovanazzi, Phys. Rev. Lett. \textbf{97}, 250402 (2006).

\bibitem{crbecexp4} T. Koch, T. Lahaye, J. Metz, B. Fröhlich, A.
Griesmaier, T. Pfau, Nature Phys. \textbf{97} 218 (2008).

\bibitem{crbecexp5} T. Lahaye, J. Metz, B. Fröhlich, T. Koch, M.
Meister, A. Griesmaier, T. Pfau, H. Saito, Y. Kawaguchi, and M. Ueda,
Phys. Rev. Lett. \textbf{101}, 080401 (2008).

\bibitem{dybecexp} M. Lu, N. Q. Burdick, S. H. Youn, and B. L. Lev,
Phys. Rev. Lett. \textbf{107}, 190401 (2011).

\bibitem{dyfermiexp} M. Lu, N. Q. Burdick, and B. L. Lev, Phys. Rev.
Lett. \textbf{108}, 215301 (2012).

\bibitem{erbecexp} K. Aikawa, A. Frisch, M. Mark, S. Baier, A. Rietzler,
R. Grimm, and F. Ferlaino, Phys. Rev. Lett. \textbf{108}, 210401 (2012).

\bibitem{erfermiexp} K. Aikawa, A. Frisch, M. Mark, S. Baier, R.
Grimm, F. Ferlaino, Phys. Rev. Lett. \textbf{112}, 010404 (2014).

\bibitem{erexp3} A. Frisch, M. Mark, K. Aikawa, F. Ferlaino, J. L.
Bohn, C. Makrides, A. Petrov, S. Kotochigova, Nature \textbf{507},
475 (2014).

\bibitem{erexp4} K. Aikawa, A. Frisch, M. Mark, S. Baier, R. Grimm,
J. L. Bohn, D. S. Jin, G. M. Bruun, F. Ferlaino, arXiv: 1405.1537.

\bibitem{erexp5} K. Aikawa, S. Baier, A. Frisch, M. Mark, C. Ravensbergen,
F. Ferlaino, arXiv: 1405.2154.

\bibitem{KRbhighdensityexp} K.-K. Ni, S. Ospelkaus, M. H. G. de Miranda,
A. Pe'er, B. Neyenhuis, J. J. Zirbel, S. Kotochigova, P. S. Julienne,
D. S. Jin, and J. Ye, Science \textbf{322}, 231 (2008).

\bibitem{KRbfermiexp} K. K. Ni, S. Ospelkaus, D. Wang, G. Quemener,
B. Neyenhuis, M. H. G. de Miranda, J. L. Bohn, J. Ye, and D. S. Jin,
Nature \textbf{464}, 1324 (2010).

\bibitem{KRbreactionexp} M. H. G. de Miranda, A. Chotia, B. Neyenhuis,
D. Wang, G. Que\'{}me\'{}ner, S. Ospelkaus, J. L. Bohn, J. Ye, and
D. S. Jin, Nature Physics \textbf{7}, 502 (2011).

\bibitem{shaperesonanceyou} B. Deb and L. You. Phys. Rev. A, \textbf{64},
022717 (2002).

\bibitem{shaperesonanceblume} K. Kanjilal and D. Blume, Phys. Rev.
A, \textbf{78} 040703(R), (2008).

\bibitem{shaperesonancehui} Z. Shi, R. Qi, and H. Zhai, Phys. Rev.
A \textbf{85}, 020702(R) (2012).

\bibitem{bohnnjp} J. L. Bohn, M. Cavagnero and C. Ticknor, New. Jour.
Phys., \textbf{11}, 055039 (2009).

\bibitem{yiandyou} S. Yi and L. You Phys. Rev. A \textbf{61}, 041604
(2000).

\bibitem{BA2} A. Derevianko, Phys. Rev. A \textbf{67}, 033607 (2003);
erratum, Phys. Rev. A \textbf{72}, 039901 (2005).

\bibitem{BA3} M. A. Baranov, M. S. Mar'enko, Val. S. Rychkov, and
G. V. Shlyapnikov, Phys. Rev. A \textbf{66}, 013606 (2002).

\bibitem{BA4} R. Liao and J. Brand, Phys. Rev. A \textbf{82}, 063624
(2010).

\bibitem{BA5} T. Shi, J.-N. Zhang, C.-P. Sun, and S. Yi, Phys. Rev.
A \textbf{82}, 033623 (2010).

\bibitem{daweiwang} D. Wang, New. Jour. Phys., \textbf{10}, 053005
(2008).

\bibitem{bohnxxx} J. L. Bohn and D. S. Jin, Phys. Rev. A \textbf{89},
022702 (2014).

\bibitem{DWBA1} A. Messiah, \emph{Quantum Mechanics}, (John Wiley
\& Sons, New York, 1958), Vol. 1.

\bibitem{DWBA2} N. R. Newbury, A. S. Barton, G. D. Cates, W. Happer,
and H. Middleton, Phys. Rev. A \textbf{48}, 4411 (1993).

\bibitem{taylor} J. R. Taylor, \textit{Scattering Theory}, Wiley,
New York, 1972.

\bibitem{rmp} Thorsten Köhler, Krzysztof Góral and Paul S. Julienne,
Rev. Mod. Phys., {\bf 78}, 1311 (2006).

\bibitem{lel} According to the effective-range theory, the $p$-wave
scattering amplitude for the ``fast-decay'' type SRI is given by
$F_{1}^{\rm (sr)}(k)=-\frac{1}{ik+\frac{1}{vk^{2}}+\frac{1}{r_{{\rm eff}}}},$
where $v$ and $r_{{\rm eff}}$ are the scattering volume and effective
range, respectively. The low-energy limit we considered in Sec. IV.
A is the limit $k\ll\sqrt{|r_{{\rm eff}}/v|}$. In this limit we have
$F_{1}^{\rm (sr)}(k)\approx-vk^{2}$. 

\bibitem{significant} Due to the facts $r_{a}<b$ and $r_{{\rm eff}}\sim b$,
the low-energy condition $k\ll\sqrt{|r_{{\rm eff}}/v|}$ is still
satisfied when $k\sim r_{a}^{2}/|v|$.

\bibitem{bohnandjin} C. Ticknor, C. A. Regal, D. S. Jin, and J. L.
Bohn, Phys. Rev. A {\bf 69}, 042712 (2004).

\bibitem{Bo Gao} B. Gao, Phys. Rev. A \textbf{58}, 4222 (1998).

\bibitem{Bo Gao 2nd} B. Gao, Phys. Rev. A \textbf{58}, 1728 (1998).\end{thebibliography}
\end{document}